%% file: acl_latex.tex
\pdfoutput=1

\documentclass[11pt]{article}

\usepackage[final]{acl}

\usepackage{times}
\usepackage{latexsym}

\usepackage[T1]{fontenc}

\usepackage[utf8]{inputenc}

\usepackage{microtype}

\usepackage{inconsolata}
\usepackage{lipsum}
\usepackage{comment}
\usepackage{multirow}

\usepackage{graphicx}
\usepackage{subcaption}
\usepackage{ragged2e}  
\usepackage{caption}   

\usepackage{url}
\usepackage{hyperref}
\usepackage{booktabs}

\usepackage{xspace}
\usepackage{enumitem}
\usepackage{tabularx}

\newcommand{\fambiental}{\textit{ReClaim}\xspace}
\newcommand{\compete}{\textit{IExp}\xspace}

\title{Leveraging LLMs to Streamline the Review of Public Funding Applications}

\author{João D.S. Marques\textsuperscript{1,2*}, André V. Duarte\textsuperscript{1,2,3*}, \\ {\bf André Carvalho\textsuperscript{2}}, {\bf Gil Rocha\textsuperscript{2}}, {\bf Bruno Martins\textsuperscript{1,2}}, {\bf Arlindo L. Oliveira\textsuperscript{1,2}}\\
         \textsuperscript{1}Instituto Superior Técnico,
         \textsuperscript{2}INESC-ID, \textsuperscript{3}Carnegie Mellon University\\
         \texttt{\{joao.p.d.s.marques, andre.v.duarte, arlindo.oliveira\}@tecnico.ulisboa.pt}\\}

\begin{document}
\maketitle
\begin{abstract}
Every year, the European Union and its member states allocate millions of euros to fund various development initiatives. However, the increasing number of applications received for these programs often creates significant bottlenecks in evaluation processes, due to limited human capacity. In this work, we detail the real-world deployment of AI-assisted evaluation within the pipeline of two government initiatives: (i) corporate applications aimed at international business expansion, and (ii) citizen reimbursement claims for investments in energy-efficient home improvements. While these two cases involve distinct evaluation procedures, our findings confirm that AI effectively enhanced processing efficiency and reduced workload across both types of applications. Specifically, in the citizen reimbursement claims initiative, our solution increased reviewer productivity by 20.1\%, while keeping a negligible false-positive rate based on our test set observations. These improvements resulted in an overall reduction of more than 2 months in the total evaluation time, illustrating the impact of AI-driven automation in large-scale evaluation workflows.

\end{abstract}

\section{Introduction}
\let\thefootnote\relax\footnote{* Equal contribution.}In the last few years, Large Language Models (LLMs) have dramatically reshaped the capabilities and expectations around automated text and image processing tasks, having shown remarkable proficiency across a wide range of domains, including translation, code generation, and mathematical reasoning, among others \cite{Kexun_Algo, Tower, deepseek_r1}.

\begin{figure}[t]
  \includegraphics[width=\columnwidth]{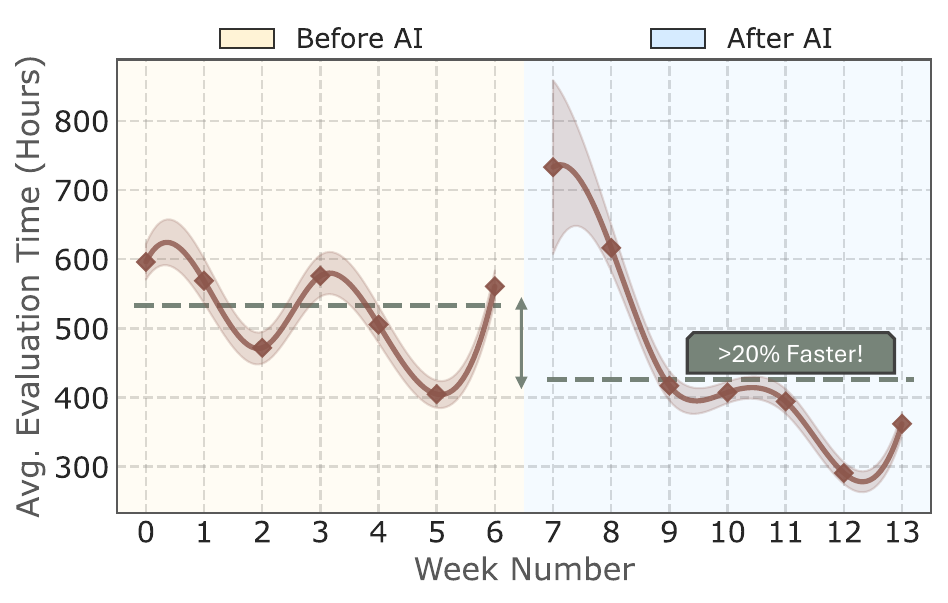}
  \vspace{-1.5em}
  \caption{Average application evaluation time within the \fambiental initiative, demonstrating a reduction of over 20\% following the deployment of our solution.}
  \label{fig:crown_figure}
\end{figure}

\par

Despite these impressive capabilities, their real-world adoption has been approached with caution due to concerns over reliability and potential misapplications \cite{EU_AI_Act_2021}. Without proper safeguards, automated systems may introduce biases, propagate misinformation, or operate with unintended autonomy, leading to unforeseen consequences \cite{Stanford_Risks,Agentic_Risks}. However, these risks can be mitigated by using LLMs as a supportive tool rather than a standalone decision-maker, ensuring that human oversight plays a key role in the process. This way, AI can be leveraged to enhance efficiency, while keeping critical processes accountable and reliable.
\par
One particular domain where automated assistance can offer tangible benefits is in public funding allocation programs across the European Union (EU). Annually, the EU and its member states allocate millions of euros through various development initiatives aimed at supporting economic growth, sustainability, and innovation \cite{Europe_2030}. These initiatives generate substantial interest, prompting thousands of individuals, businesses, and institutions to submit applications. Unfortunately, this high volume of submissions often surpasses the current processing capacities of human evaluation teams, leading to significant bottlenecks and delays in application processing \cite{DECO_Critica}. Such delays translate into prolonged evaluation times, widespread dissatisfaction, and ultimately a deterioration of public confidence in the efficiency of these initiatives \cite{Expresso_Critica}.
\par
To address these challenges and explore the potential for AI-driven automation in public funding allocation, we introduce two distinct LLM-based systems designed to enhance human-in-the-loop evaluation processes. These were deployed in two EU-funded initiatives in Portugal: (i) corporate applications aimed at international business expansion \cite{Compete}  and (ii) citizen reimbursement claims for investments in energy-efficient home improvements \cite{Fundo_Ambiental}.
\par
In the first initiative, the main focus is on producing high-quality summaries of applications - the task reviewers identified as the biggest bottleneck. With each submission averaging 30,000 tokens (over 50 pages), manual summarization is both time-consuming and resource-intensive.
\par 
In the second initiative, the objective is to ensure consistency between claimed expenses and the supporting documents submitted by applicants. These documents contain unstructured information, requiring a combination of classical Optical Character Recognition (OCR) techniques and LLMs to extract and structure key details. This enables the automatic pre-filling of a verification checklist with over 80 mandatory review items per application. While the task itself is highly deterministic, due to strict submission guidelines, the real challenge lies in its scale: processing around 80,000 applications, each averaging eleven user-uploaded documents (totaling $\approx$ 880,000 documents) that would otherwise require full manual review.

\par
At the time of writing, our systems have been deployed in the field for over three months, demonstrating both quantitative and qualitative improvements. We find the most significant gains to be in the reimbursement claims initiative, where reviewer productivity increased by $\approx$20\% (Figure~\ref{fig:crown_figure}), highlighting that structured document verification tasks are particularly well-suited for automation.
\par

\par  

Our main contributions are as follows:  
\begin{itemize}  
    \item We report on the successful deployment of two AI-assisted document evaluation systems, demonstrating how automation can accelerate application analysis while ensuring human oversight for decision-making integrity.
    \item We provide a discussion on the key lessons learned from the real-world deployment, offering best practices for integrating AI models into similar settings to ours.
\end{itemize}

\section{Related Work}

Automating document review has long been a central goal across both the public and private sectors. Today, the landscape is rapidly changing, largely due to the transformative capabilities of LLMs \cite{GPT-4o, deepseek_r1}. These models offer unprecedented flexibility, adaptability to diverse document types, and the ability to generalize across a wide range of unstructured data formats \cite{adapted_LLMS,lumberchunker}. However, it is important to recognize that automated document processing predates the arrival of LLMs. Traditionally, such automation relied on rule-driven natural language processing (NLP) and optical character recognition (OCR) techniques \cite{classical_ocr}. While these systems could achieve good performance in controlled environments, they were highly sensitive to variations in document structure and content, which made them difficult to maintain and challenging to extend to broader applications \cite{rule_based_ie_is_dead, rule_based_ie}.
\par
The shift from traditional to LLM-driven approaches is now evident across multiple domains. In the legal sector, LLM-based tools enable rapid review and extraction from a wide range of legal texts, streamlining processes that once required extensive manual effort \cite{lawllm}. In human resources, resume screening has evolved from basic keyword analysis \cite{cv_analysis1} to sophisticated LLM-powered systems capable of nuanced candidate-role matching \cite{cv_analysis2}. Public administration \cite{EU2024_AI_stratigic} offers a further illustrative example: while the European Commission’s 2020 AI Watch report \cite{AI_Watch_Report} documented widespread adoption of conventional machine learning solutions within government applications, the Commission’s 2024 AI@EC strategic vision report  \cite{AI@EC} signals a clear commitment to generative AI and the integration of LLMs into these processes.
\par
Despite notable successes, the adoption of LLMs in these sectors has also exposed significant challenges. Failures due to bias, lack of transparency, or over-reliance on unsupervised automation have led to widely publicized setbacks \cite{amazon_bias,machine_learning_bias}. These experiences have shaped the current best practices for LLM deployments, with most organizations now embedding explicit human-in-the-loop review at key decision points, to ensure oversight and maintain accountability of the systems \cite{human_oversight}.
\par 
Our work follows precisely this trajectory by reporting on two instantiations of LLM-based systems developed for Portuguese government initiatives. In both cases, efficiency gains are balanced with continuous human oversight, supporting trustworthy automation in complex and socially significant document evaluation workflows.

\section{AI-Assisted Evaluation: Overview of Initiatives and Proposed Pipelines}

\begin{figure*}[ht]
\begin{minipage}[t]{.35\textwidth}
\centering
\includegraphics[width=0.8\textwidth]{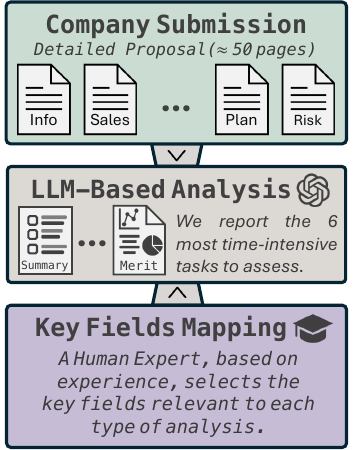}
\caption{The \compete review system leverages GPT-4o to automate the six most time-consuming tasks. Before the analysis, reviewers segment and filter the proposal to avoid overloading the LLM with irrelevant context.}
\label{fig:compete_pipeline}
\end{minipage}
\hfill
\begin{minipage}[t]{.6\textwidth}
\centering
\includegraphics[width=1\textwidth]{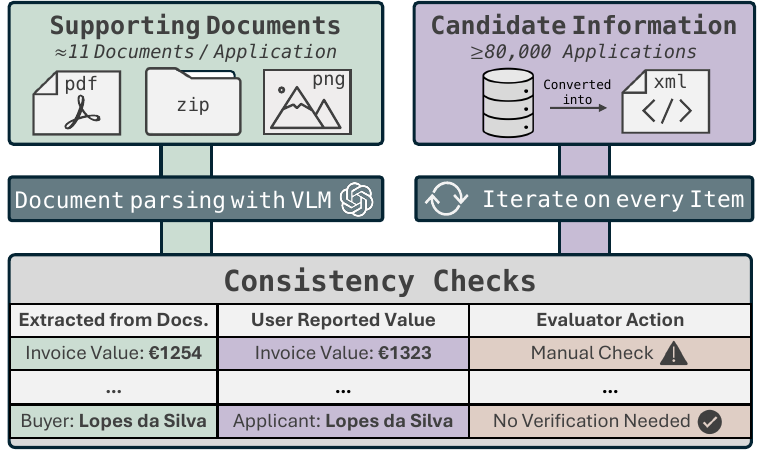}
\caption{The \fambiental evaluation system processes a large volume of supporting documents submitted by citizens in various formats. These documents are automatically parsed using GPT-4o, which extracts critical details and performs automated consistency checks, flagging discrepancies for manual reviewer verification.}
\label{fig:FA_pipeline}
\end{minipage}
\end{figure*}

Both instantiations aim to streamline the evaluation of large-scale funding programs, although each comes with distinct requirements, prompting us to develop custom solutions for each context. Here, we summarize the main objectives of the two programs, followed by an explanation of how we adapted our systems to meet their particular demands.

\subsection{Corporate Applications for International Expansion (IExp)}
\label{ssec:iexp_pipeline_overview}

The \compete initiative \cite{Compete} aims to strengthen the international competitiveness of Portuguese Small and Medium Enterprises (SMEs). With a total allocation of 32 million euros, this program invites companies to apply for funding to support projects specifically aimed at expanding their business models and integrating more effectively into global value chains.
\par
To qualify for funding, companies must submit a comprehensive application form detailing their business plans, internationalization strategies, historical and projected market analysis, and associated risks, which results in documents spanning more than 50 pages per submission.
\par
To determine the overall eligibility of the application, reviewers complete a comprehensive evaluation process that involves verifying supporting documentation, cross-checking key information against external government databases, extracting and summarizing relevant details, and assigning scores across multiple criteria. While applications are ultimately reviewed with care and precision, the complexity and thoroughness of the evaluation process leads to extended timelines and a significant burden on reviewers.
\par
Our AI-assisted system was designed to address the most time-consuming and objective elements of the process. Through ongoing discussions with the evaluation team, we prioritized the summarization of the application, assisting in detecting internal inconsistencies across documents and in assigning preliminary scores and justifications for selected sections. In total, our approach automates six specific tasks within the reviewer workflow, detailed in Appendix \ref{app:full_compete}.
\par
As illustrated in Figure \ref{fig:compete_pipeline}, our pipeline leverages GPT-4o \cite{GPT-4o}, which is prompted with the same guidelines the human evaluators follow. For each task, the LLM receives the most relevant sections of the application as the only input. The most relevant sections for each task were identified by the evaluation team, who contributed with their domain expertise by sharing this information. This targeted input not only helps to address the LLM's `lost in the middle' problem \cite{lost_in_the_middle}, but also makes the prompt more cost-efficient.

\subsection{Reimbursement Claims for Energy-Efficient Home Investments (ReClaim)}
\label{ssec:reclaim_pipeline_overview}

The \fambiental initiative \cite{Fundo_Ambiental} is part of Portugal's \textit{More Sustainable Buildings} program, and has the core objective of improving the energy and environmental performance of Portuguese residential buildings. With a total budget of 30 million euros, the program covers a diverse range of intervention categories, such as window replacement, thermal insulation, HVAC system upgrades, solar installations, and water efficiency measures. Applicants can submit multiple applications across these categories, but each application must focus on a single typology and sub-typology, as detailed in Appendix~\ref{app_full_fa}.
\par
Unlike in the case of the \compete initiative, the \fambiental process is fundamentally a large-scale verification task. The program received approximately 80,000 applications, each accompanied by a bundle of supporting documents (eleven documents per application on average). The main challenge was not the need for subjective interpretation, but rather managing the substantial variability in how documents were submitted. Documents were provided in a range of formats, essential documents were sometimes missing or incorrectly categorized, and the information reported by applicants did not always correspond to the details found in invoices or official declarations.
\par
To handle this variability, we first standardized the scope of data processed by our solution. The pipeline supports only widely used file types, such as PDF, ZIP, and PNG. Roughly 10\% of documents fell outside this criterion and were automatically excluded from automated parsing. For each excluded document, the reviewer received a notification explicitly indicating the presence of unsupported data and the need for manual verification.
\par
The core of the \fambiental solution is then a hybrid processing pipeline (Figure~\ref{fig:FA_pipeline}) that combines classical document parsing and manipulation with VLM-driven information extraction. First, all user-provided form fields are converted into a structured XML format. Then, every supporting document is mapped to its corresponding application. GPT-4o is then used to parse unstructured supporting documents, extracting key details such as invoice values, buyer information, and intervention descriptions. In the final step, the system conducts automated consistency checks by comparing the extracted information against the values reported by the applicant. When a match is found, the item is marked as ``No Verification Needed," allowing reviewers to progress efficiently through the checklist. In cases where discrepancies arise (e.g., if the extracted invoice value differs from the user-reported value), the system flags the item for ``Manual Check," thereby directing the reviewer’s attention to items that require targeted intervention.

\section{System Design \& Implementation}

The deployment of automated systems in real-world settings requires careful consideration of three main aspects: ensuring safety and security, balancing the cost-performance tradeoff, and performing a smooth integration with existing workflows.

\subsection{Safety and Security}

To comply with the \textit{General Data Protection Regulation} (GDPR) \cite{GDPR}, our systems are designed to keep the data within the EU borders at all times. Processing occurs exclusively within the region, and storage is managed on encrypted local disks with restricted access, safeguarding confidentiality and integrity.
\par
Beyond data protection, we ensure the reliability of our systems through a human-in-the-loop design. This means that our outputs serve strictly as recommendations, ensuring human reviewers retain oversight and accountability at all times.

\subsection{Balancing Cost and Performance}

Deploying AI-powered systems at scale requires carefully balancing the available budget with the expected performance, making model selection a critical factor in the process. Given that a significant portion of our tasks consists of Visual Question-Answering (VQA), we initially considered locally hosted open-source models, which offer advantages such as faster inference times and lower operational costs. However, despite promising benchmark results in VQA tasks \cite{pixtral}, we found that VLMs like Qwen2-VL \cite{qwen2} or LLaMa-3.2 \cite{llama_3.2} perform worse in Portuguese, the main language of our data.
\par
Following discussions with the evaluation teams, we prioritized minimizing false negatives, particularly in the reimbursement claims task, where such errors significantly increase the manual verification workload. Given that cost was not a major limiting factor, we opted to use top-performing closed-source models. For that, we conducted a blind test where evaluators assessed anonymized outputs from GPT-4o \cite{GPT-4o} and Gemini-1.5 Pro \cite{gemini_1.5}. Evaluators consistently preferred GPT-4o, leading to its selection for both initiatives. Further details are given in Appendix~\ref{sec:blind_test}.

\subsection{Integration with Existing Workflows}

Adapting our solution into the existing workflow of the \compete initiative was relatively straightforward, as our method was developed in parallel with the software used by the reviewers. This early involvement ensured a smooth transition, allowing evaluation procedures to be designed with automated assistance in mind.
\par
The \fambiental initiative, however, presented more challenges, because the evaluation process was already in progress when we joined. Reviewers were not used to working with AI-generated outputs, so we had to refine the system step by step, working closely with them to ensure compliance with their needs. Another difficulty was that the evaluation platform is not owned by the evaluation team but instead managed by an external provider. This limited how much we could integrate our functionalities directly. Despite these constraints, continuous feedback helped us improve the system over time, increasing reviewers' confidence in the solution and making the process more efficient.

\section{Results}

We evaluated our systems on both quantitative improvements and reviewer feedback.

\begin{figure}[t]

  \includegraphics[width=\columnwidth, trim=10 10 10 10, clip]{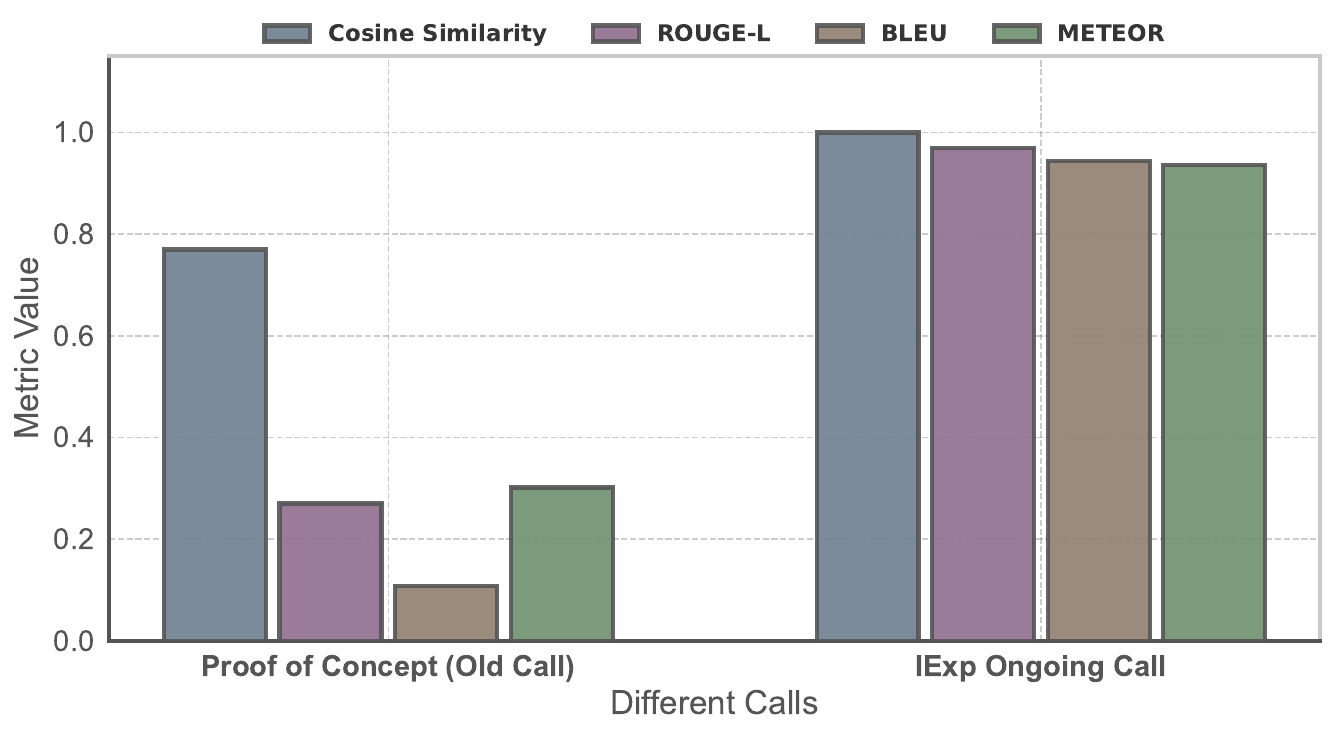}
  
  \caption{Agreement between application summary during the POC and the current application call.}
  \label{fig:poc_vs_now}
\end{figure}

\subsection{Quantitative Improvements}

\textbf{Corporate Applications (\compete):} We focus our quantitative analysis on two key metrics: the alignment between AI-generated and reviewer summaries and the classification alignment of company activity types.

\begin{figure}[h!]
\centering
  \includegraphics[width=0.9\columnwidth]{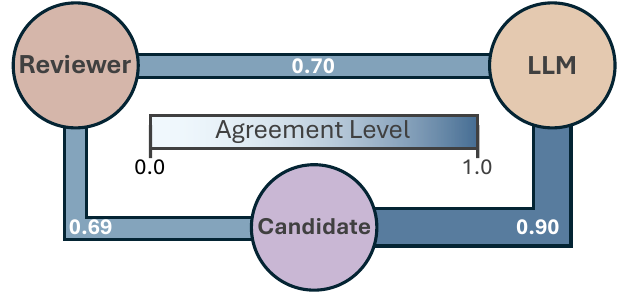}
  \caption{Agreement on classifying activities as either marketing or organizational, between applicants, reviewers, and the LLM.}
  \label{fig:agreement_compete_atividades}
\end{figure}

Summary alignment was evaluated in two settings: (a) during the proof of concept (POC), using a test set containing 50 applications from past calls reviewed without AI assistance, and (b) in the ongoing evaluation of the most recent call, using a test set containing 11 applications where reviewers are supported by our tool. As shown in Figure~\ref{fig:poc_vs_now}, the average cosine similarity improved significantly, rising from 0.77 to 0.99 between the two calls. In parallel, all other metrics (ROUGE-L, BLEU, and METEOR) also showed big improvements, increasing from below 0.35 to above 0.9. This demonstrates a substantial gain in alignment quality across all evaluation dimensions, demonstrating that the tool now effectively helps reviewers standardize evaluations, increasing speed and focus on key details. 
In the case of activity classification (i.e., identifying valid activities and labeling them as either marketing or organizational), we used a test set of 764 applications from past calls that were reviewed without AI assistance. As shown in Figure~\ref{fig:agreement_compete_atividades}, although LLMs tend to agree more with candidates, agreement with reviewers remains at around 70\%. Additional results are provided in Appendix~\ref{app:full_compete}.

\textbf{Reimbursement Claims (\fambiental):} As shown in Figure~\ref{fig:crown_figure}, reviewer productivity increased by $\approx$~20\%. An evaluation of 200 test samples, distributed evenly across typologies, demonstrated that the LLM-assisted system enabled reviewers to skip manual validation in approximately 76\% of field verifications, especially for standardized typologies. Most outputs were accurate (88\%), with the remaining errors being either minor or easily identifiable. Critical issues, such as false positives or reading errors, were rare. Detailed results can be found in Appendix~\ref{app_full_fa}.
\par
Beyond direct improvements in efficiency, we further analyzed whether the deployment of our system affected evaluator behavior and applicant responses. Specifically, we examined (1) the number of clarification requests that evaluators sent to applicants, and (2) the proportion of appeal requests submitted by applicants following a decision. As shown in Table~\ref{tab:clarifications}, both measures decreased after the adoption of AI assistance. This suggests that evaluators made fewer human-level mistakes and achieved greater consistency during their assessments, leading to fewer appeals and, consequently, indicating a more positive end-user experience.

\begin{table}[h!]
\centering
\resizebox{\columnwidth}{!}{%
\begin{tabular}{lcc}
\toprule
\textbf{Metric} & \textbf{Before AI} & \textbf{After AI} \\
\midrule
Clarification Requests / Application & 2.13 & 2.05 \\
Applicant Appeal Rate (\%) & 25.8 & 20.4 \\
\bottomrule
\end{tabular}
}
\caption{Evaluator clarification requests and applicant appeal rates before and after the solution deployment.}
\label{tab:clarifications}
\end{table}

\subsection{Reviewer Feedback}

For both tasks, reviewers were asked for feedback on our solutions. In the \compete task, evaluators estimated that AI assistance could accelerate the review process by up to 30\%, with the greatest benefit observed in generating application summaries. Secondly, and though not explicitly stated, it can be inferred that the transition toward more standardized, LLM-generated summaries has been positively received by the evaluation team.


In the case of the \fambiental task, feedback was more mixed and divided into two main groups. The first group, composed of reviewers who collaborated on the development of the AI tool, expressed strong appreciation for its usefulness. The second group included reviewers who began using the tool after its deployment. Among these, those with extensive evaluation experience often reported it as highly useful, with estimated time savings of up to 40\%. However, other reviewers either struggled to understand how to effectively use the tool, or lost confidence after encountering errors. These errors were typically minor or false alerts, but still impacted trust in the system.


Overall, the findings suggest that LLMs are already mature enough to significantly support the application review process. However, their effectiveness is highly dependent on the surrounding ecosystem, including bureaucratic context, reviewer tooling, and the stability of evaluation criteria.

\section{Conclusions}

This work explored the potential of LLMs to improve evaluation of public funding applications. Our deployments demonstrated that, when properly integrated into human-in-the-loop pipelines, LLMs can substantially accelerate evaluation workflows, reduce manual workload, and promote greater uniformity in the reviewer outputs.
\par
To conclude, we now summarize some of the key lessons learned from this project. We believe that these insights can be valuable for future developments that aim to explore the potential of LLMs in similar scenarios.

\subsection*{Organizational and Regulatory Barriers}
Although technical challenges were expected, our experience revealed that bureaucracy is often the main source of delays and reduced solution quality. Critical roadblocks included third-party platform ownership, which restricted our ability to implement system modifications; strict GDPR requirements, which narrowed the pool of viable models; and complex, multi-step authorization workflows that delayed data access. As a result, we believe that successful deployment in this domain requires not only technical flexibility but also careful planning for this constraints, which can fundamentally restrict the available technical options.

\subsection*{Polarized Adoption Patterns}
Integrating AI in a human-in-the-loop pipeline is only effective if human reviewers actually decide to use the AI during their evaluations. In our deployment, we observed that reviewers often split into two groups: those who were willing to use the tool and focus on its benefits, and those who became very cautious or critical whenever the system made a mistake. When reviewers lose trust in the system, they may stop using the AI altogether; hence, the potential gains in efficiency are not fully achieved. As a result, we found it is important to give extra attention to reviewers who are less open to using AI. By clearly explaining what the system can and cannot do, it is possible to set realistic expectations and help all reviewers become more tolerant of small system errors, leading to better overall adoption. In practice, effective change management across multiple levels of relevant processes and stakeholders is essential for successful implementation.

\par
\subsection*{High Practical Application Potential}

Perhaps the most significant takeaway is the high practical application potential of AI-assisted evaluation systems at scale. While the initial development and adaptation of each solution requires careful planning and close collaboration with the evaluation teams, the large-scale deployment is incomparably faster than manual evaluation. For example, our \fambiental system processed the $\approx$ 80,000 applications in less than three weeks. Obviously, this only impacts part of the total evaluation time, since the human-in-the-loop setting still requires significant manual work across other tasks. However, as models continue to improve, it becomes increasingly plausible that fully automated evaluation could become viable, in particular when the underlying processes are reengineered. If and when that point is reached, we can expect even more dramatic speedups in public sector workflows.


\section*{Limitations}

In the \compete initiative, the system was limited by the fact that certain reviewer tasks depended on past applications or external databases, which were not accessible to the LLM. Instead, the LLM is restricted exclusively to relevant sections of the current application (as detailed in Section \ref{ssec:iexp_pipeline_overview}). Reviewers were informed of this limitation, which influenced task selection and led us to prioritize those that required little or no external context.

In the \fambiental initiative, the system faced challenges due to inconsistent document formats and low-quality file submissions. The permissive nature of the submission platform resulted in cases where the system could not be applied. As a result, we recommended enforcing stricter file format requirements in future calls. Moreover, poor-quality content (e.g., blurry scans) negatively affected performance, emphasizing the importance of more precise applicant guidelines and implementing file verification during submission to enhance system effectiveness.

\section*{Acknowledgments}
We thank the anonymous reviewers for their valuable comments and suggestions. This research was supported by the Portuguese Recovery and Resilience Plan through project C645008882-00000055 (i.e., the Center For Responsible AI), and by Fundação para a Ciência e Tecnologia, I.P. (FCT) through the projects with references UID/50021/2025 and UID/PRR/50021/2025.

\bibliography{custom}

\newpage
\onecolumn
\appendix

\input{appendix.tex}

\end{document}

%% file: appendix.tex
\section{Model Selection - Blind Test}
\label{sec:blind_test}

To better understand which language model could align more closely with human preferences for our tasks, we conducted a blind evaluation of outputs from two language models: Gemini-1.5 Pro and GPT-4o. For 10 applications from the Corporate Applications (\compete) initiative, both models generated the application summary. Each pair of summaries was then blindly evaluated by three reviewers, who were shown the summaries in randomized order and asked to select the one they preferred.
\par
As shown in Table~\ref{tab:human_eval}, GPT-4o was preferred in 8 out of 10 cases, receiving 21 out of the 30 total votes. Based on this trend, we adopted GPT-4o for our solution, and used it across all tasks to ensure consistency.

\begin{table}[h!]
\centering
\begin{tabular}{@{}c c c l@{}}
\toprule
\textbf{Summary ID} & \textbf{GPT-4o Votes} & \textbf{Gemini Votes} & \textbf{Majority Winner} \\
\midrule
1  & 3 & 0 & GPT-4o \\
2  & 2 & 1 & GPT-4o \\
3  & 1 & 2 & Gemini  \\
4  & 2 & 1 & GPT-4o \\
5  & 3 & 0 & GPT-4o \\
6  & 2 & 1 & GPT-4o \\
7  & 2 & 1 & GPT-4o \\
8  & 1 & 2 & Gemini  \\
9  & 3 & 0 & GPT-4o \\
10 & 2 & 1 & GPT-4o \\
\midrule
\multicolumn{1}{l}{\textbf{Total Votes}}  & \textbf{21} & \textbf{9} & \textbf{GPT-4o} \\
\multicolumn{1}{l}{\textbf{Summary Wins}} & \textbf{8}  & \textbf{2} & \textbf{GPT-4o} \\
\bottomrule
\end{tabular}
\caption{Human preferences between summaries generated by GPT-4o and Gemini. Each pair of summaries (one from each model) was evaluated by three annotators.}
\label{tab:human_eval}
\end{table}

\newpage

\section{\fambiental Details}
\label{app_full_fa}

\begin{table}[h!]
\centering
\begin{tabular}{@{}c c c c c@{}}
\toprule
\textbf{ID} & \textbf{Typology} & \textbf{\# Sub-Typologies} & \textbf{Applications (\%)} & \textbf{Avg Documents} \\
\midrule
1 & Window Replacement & None & 27.34\% & 14 \\
2 & Thermal Insulation & 4    & 1.36\% & 12 \\
3 & Heating and Cooling Systems & 3    & 48.06\% & 10 \\
4 & Solar Panels & 2    & 32.72\% & 11  \\
5 & Water Efficiency & None    & 0.52\% & 8  \\
\bottomrule
\end{tabular}
\caption{Typology distribution for more than 80,000 \fambiental applications that were analyzed.}
\label{tab:typologies}
\end{table}

\begin{figure}[h!]
  \centering\includegraphics[width=0.7\columnwidth, trim=50 0 50 0, clip]{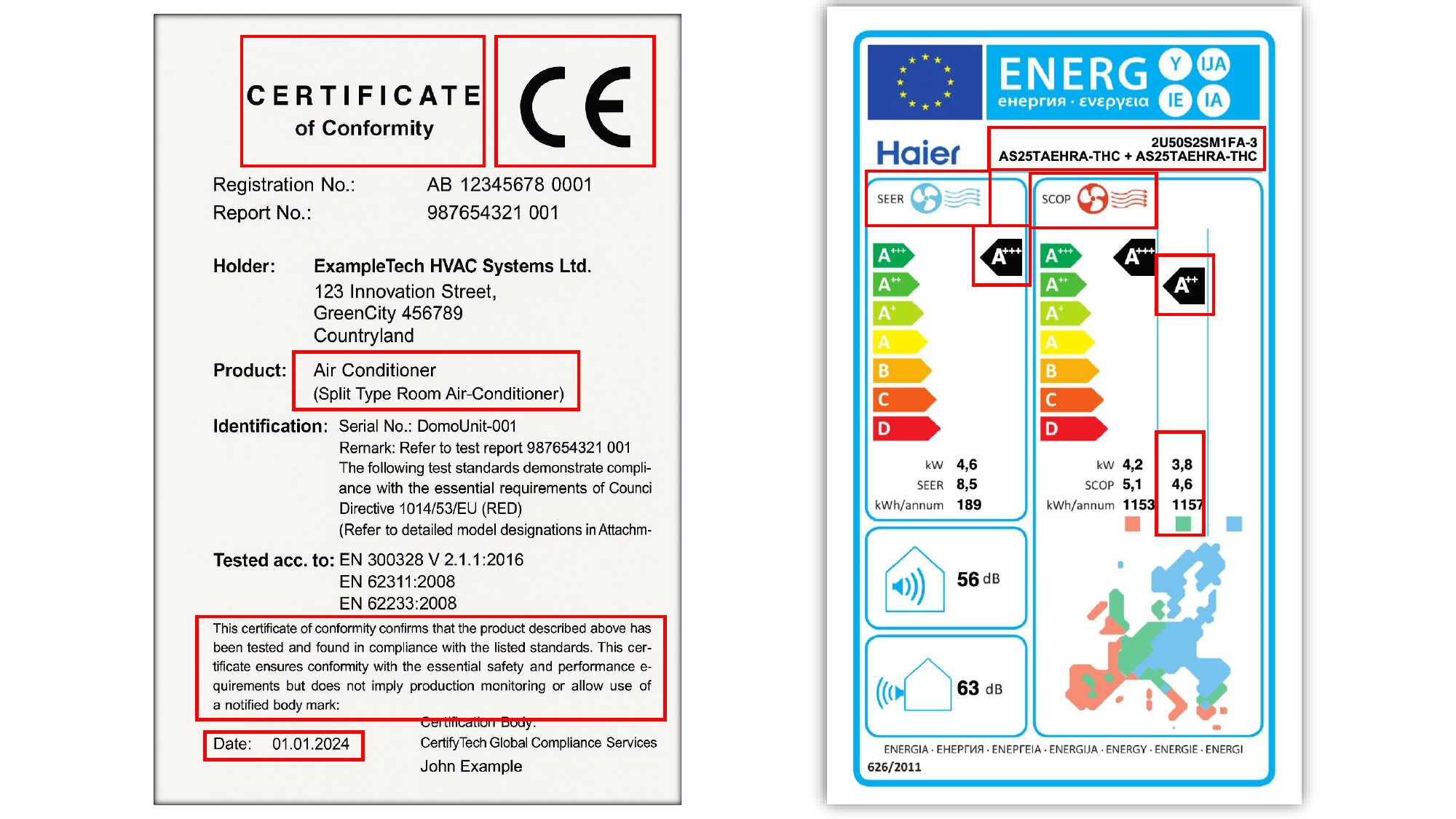}
  \caption{\fambiental application files examples.}
  \label{fig:fa_files}
\end{figure}

\begin{table}[h!]
\centering
\begin{tabularx}{0.7\textwidth}{X >{\centering\arraybackslash}X >{\centering\arraybackslash}X}
\toprule
\textbf{Report} & \textbf{Cost (€)} & \textbf{Time (s)} \\
\midrule
Typology 1       & 0.05 & 37 \\
Typology 2.1.1   & 0.06 & 61 \\
Typology 2.1.2   & 0.02 & 34 \\
Typology 2.2.1   & 0.02 & 24 \\
Typology 2.2.2   & 0.09 & 108 \\
Typology 3.1     & 0.02 & 41 \\
Typology 3.2     & 0.10 & 87 \\
Typology 3.3     & 0.09 & 23 \\
Typology 4       & 0.04 & 39 \\
Typology 5.1     & 0.03 & 25 \\
Typology 5.2     & 0.21 & 173 \\
\midrule
All Typologies Avg.   & 0.06 & 48 \\
Eligibility      & 0.01 & 13 \\
Common Core      & 0.02 & 29 \\
\midrule
\textbf{Total}   & \textbf{0.09} & \textbf{90} \\
\bottomrule
\end{tabularx}
\caption{Cost and time analysis for \fambiental applications on average.}
\label{tab:time_cost_fa}
\end{table}

\newpage 

\subsection{Additional Results}

\begin{figure}[h!]
  \centering\includegraphics[width=0.7\columnwidth, trim=0 0 0 0, clip]{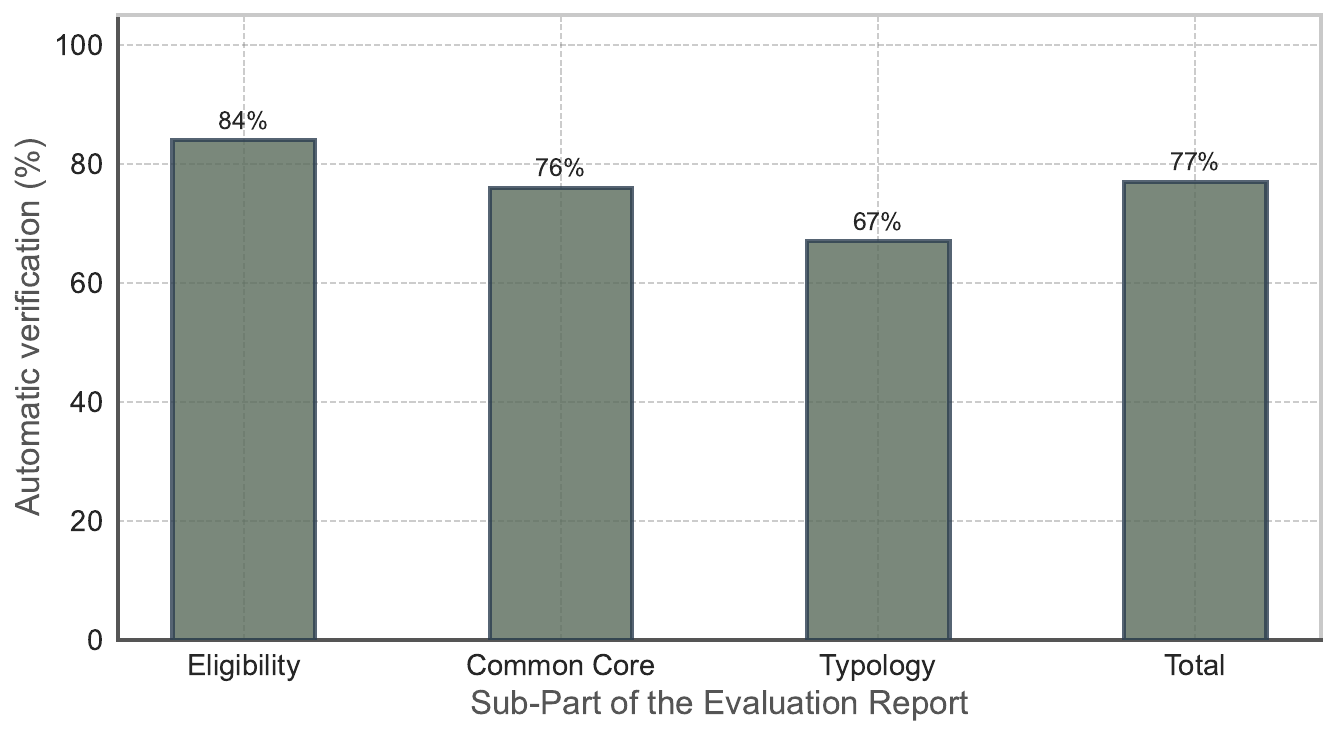}
  \caption{Proportion of field verifications that can suppress manual validation across the main sections of the application. A verification is considered not to require manual review if it is either correct or clearly incorrect in an interpretable way (e.g., a misplaced digit or incorrect capitalization). On average, 76\% of verifications did not require human intervention.}
  \label{fig:fa_validation_need}
\end{figure}

\begin{figure}[h!]
  \centering\includegraphics[width=0.72\columnwidth, trim=0 0 0 0, clip]{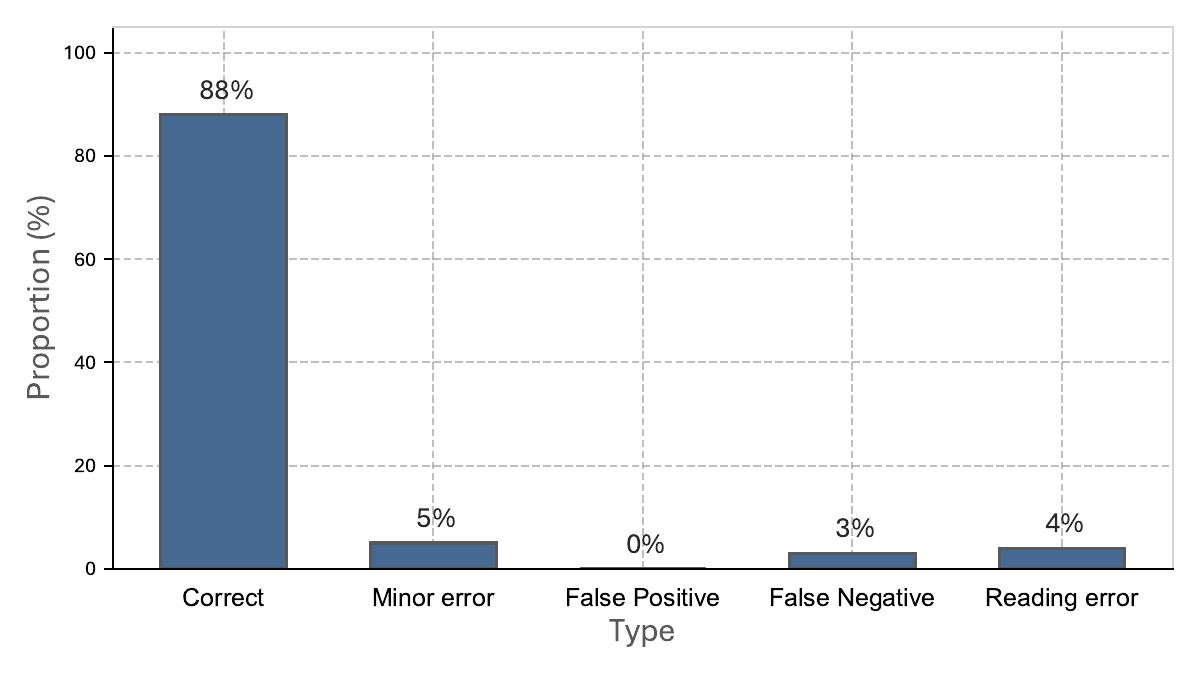}
  \caption{Evaluation of the correctness of $\approx$ 7,000 verification fields obtained from 200 applications. This analysis categorizes the types of errors produced by the LLM-generated reports. Minor errors refer to mistakes that evaluators can immediately recognize and correct, such as misspellings, capitalization issues, or false alerts. False positives occur when the system fails to detect an actual error. False negatives are instances where the system incorrectly flags an error when there is none. Reading errors result from the LLM being unable to read or process a file properly.}
  \label{fig:fa_errors}
\end{figure}

\newpage

\subsection{\fambiental Prompt Example}
\label{app:prompts_fa}

\begin{table}[h!]
  \centering
  \begin{tabularx}{0.95\textwidth}{X}
    \toprule[1.1pt]

    \textbf{System Prompt:} You will receive a document as an attachment (MCP). Please help convert part of it into XML. The XML tags to be created are:\\ \\

    <mcp\_type> (You must classify the type of document. The main categories are: \\
    1. Submission Receipt – issued by DGEG \\
    2. Screenshot of the MCP platform submission \\
    3. Confirmation email of MCP submission \\
    4. Document recognizing technician or company responsible for private electrical installations \\
    5. Document granting exemption from prior control) \\ \\

    <ID\_energy\_producer> (ID of the energy producer)\\

    <NIF\_NIPC\_mcp> (Tax ID of the energy producer) \\

    <address\_mcp> (Address of the installation) \\

    <energy\_source\_mcp> (Source of energy)

    \par

    <generator\_power\_mcp> (Nominal power of installed generators. Include units when possible)

    \par

    <nominal\_power\_mcp> (Installed capacity/nominal power of the inverter. Include units when possible)

    \par

    <date\_start\_mcp> (Date of authorization to begin operation)

    \par

    <date\_submission\_mcp> (Date of MCP submission)

    \par \\

    If any of the values are missing in the document, return the corresponding tag with 'None'.

    \par

    All responses should follow the format: <tag\_name>value</tag\_name>

    \par

    If a document doesn't match any listed type, use: <mcp\_type>None</mcp\_type>

    \\
    \bottomrule[1.1pt]
\end{tabularx}
  \caption{Example of the system prompt used for the Prior Communication (MCP) document-to-XML task. The user prompt consists of a set of images corresponding to the attached document. Each required tag is extracted using structured JSON query formatting, as shown in the schema. Values are returned in XML-style tags (e.g., <tag\_name>value</tag\_name>) and are defined through a JSON schema that specifies their type, description, and possible enumerations. Most prompts for similar tasks follow this same structure, varying only in the specific tags and extraction goals. Once extracted, the data is filtered and either passed to subsequent LLM calls or used in downstream logic (e.g., checking if one date is later than another).}
\end{table}

\newpage

\subsection{\fambiental Reports Examples}
\label{sec:appendix_fa_eligibility_report}

\subsubsection{Eligibility Report Example}

\begin{figure}[h!]
  \fbox{
  \includegraphics[width=\columnwidth, trim=50 220 50 60, clip]{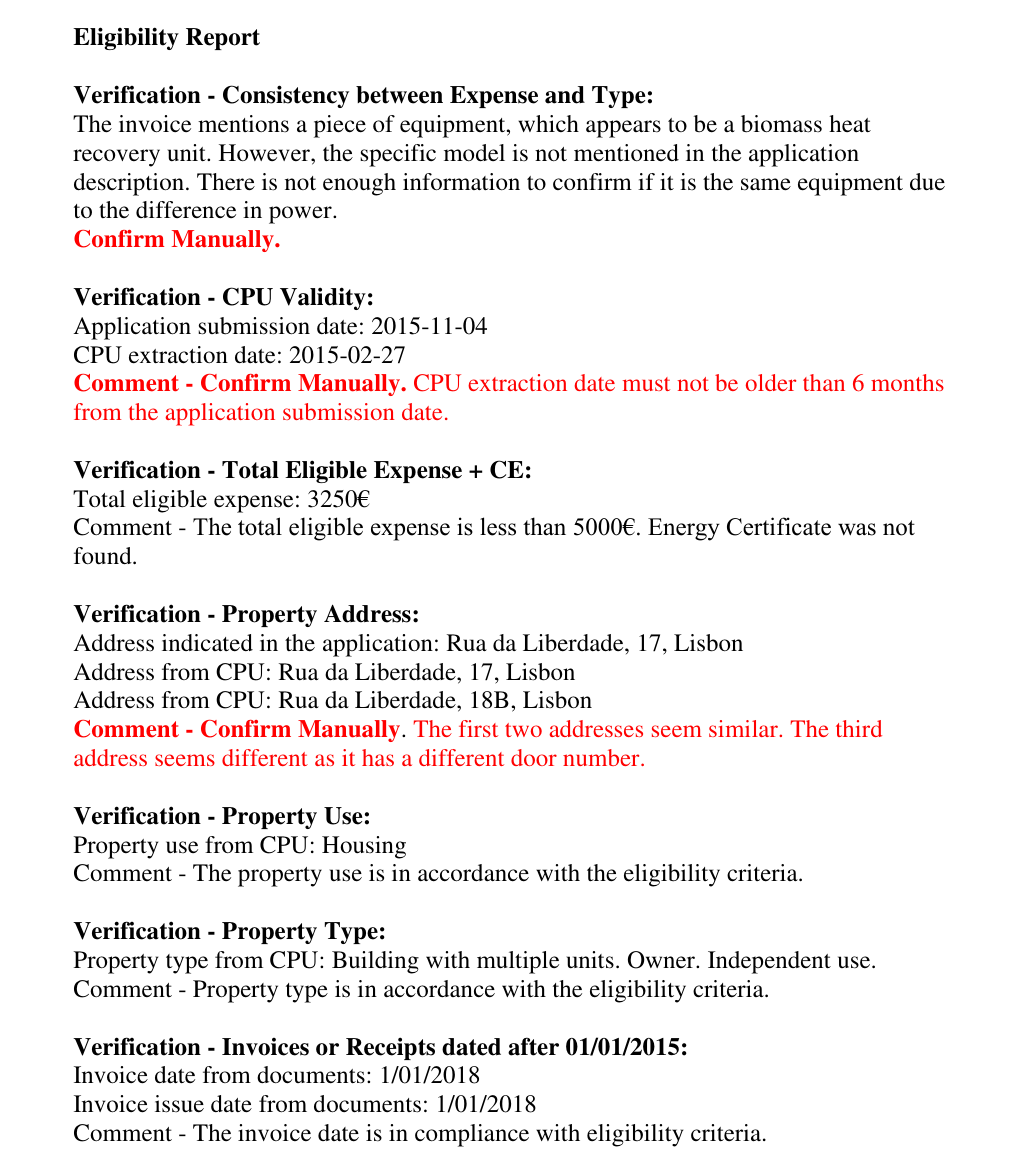}
  }
  \caption{Eligibility report dummy example.}
  \label{fig:elig_report}
\end{figure}

\newpage
\subsubsection{Common Core Report Example}
\begin{figure}[h!]
  \fbox{
  \includegraphics[width=\columnwidth, trim=50 100 50 60, clip]{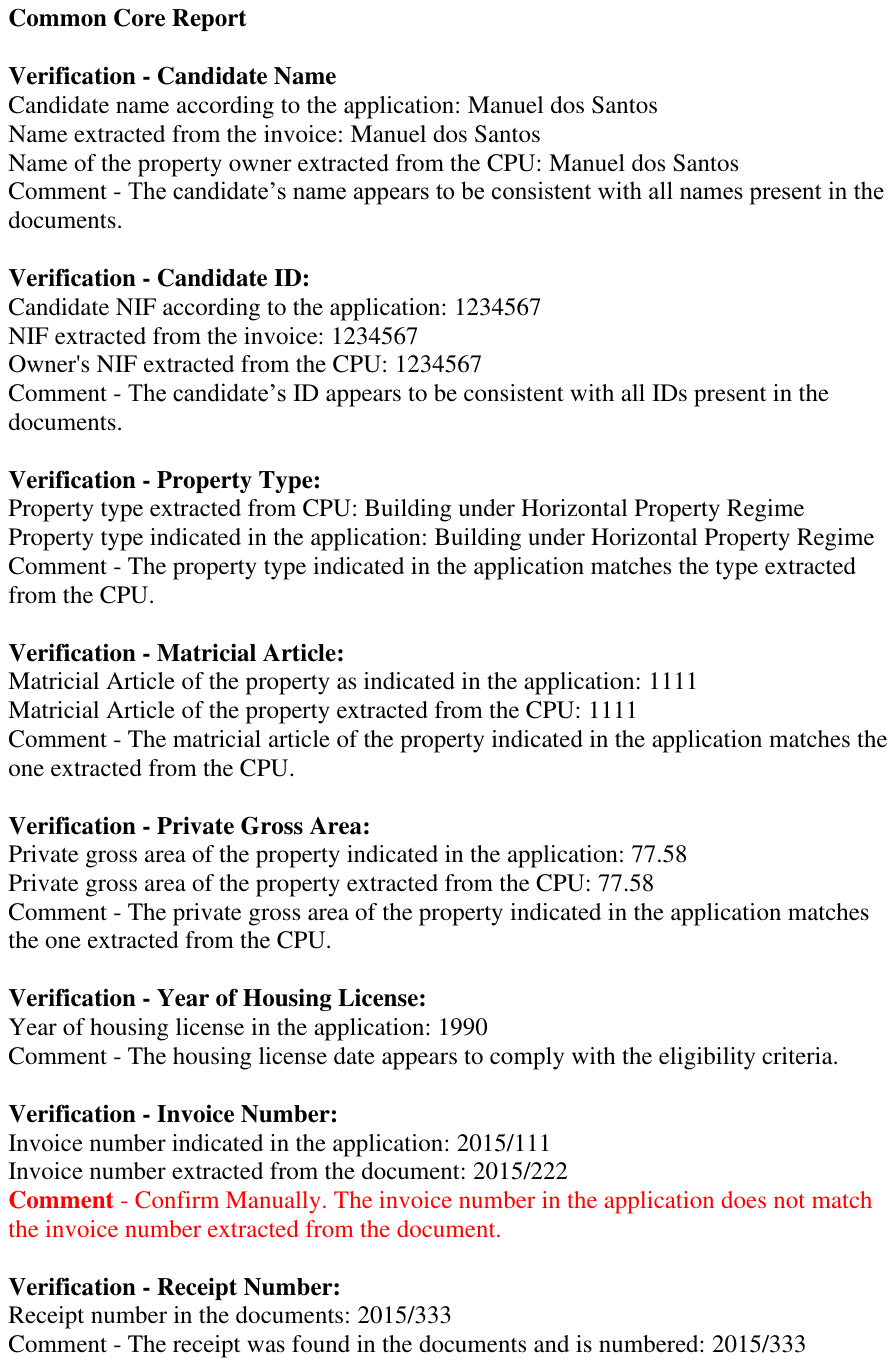}
  }
  \caption{Common core report dummy example.}
  \label{fig:com_report}
\end{figure}

\newpage
\subsubsection{Typology Report Example}
\begin{figure}[h!]
  \fbox{
  \includegraphics[width=\columnwidth, trim=50 100 50 60, clip]{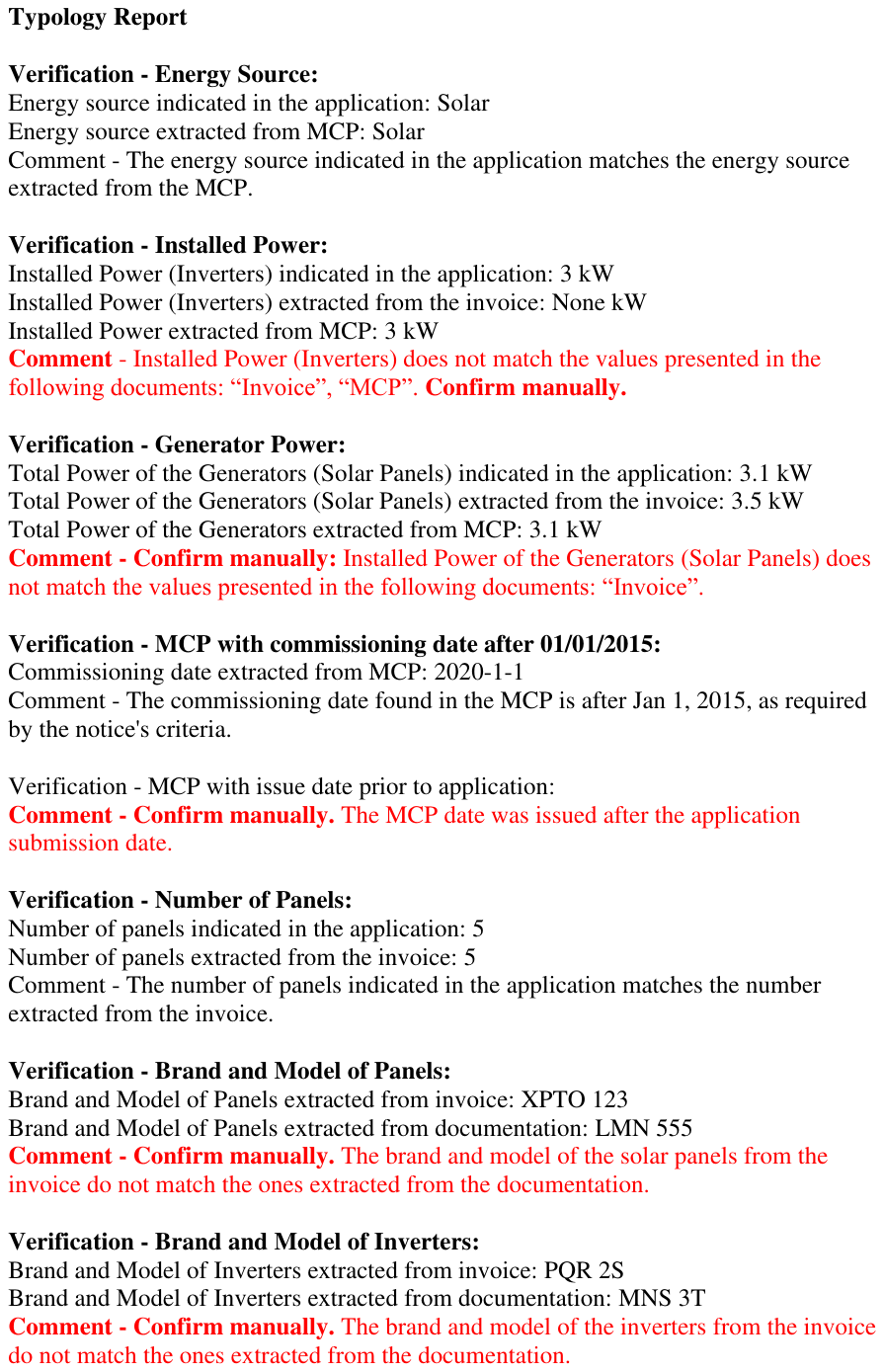}
  }
  \caption{Typology report dummy example.}
  \label{fig:tip_report}
\end{figure}

\newpage 

\section{\compete Details}
\label{app:full_compete}

\begin{table}[h!]
\centering
\begin{tabularx}{0.9\textwidth}{
  >{\centering\arraybackslash}X
  >{\centering\arraybackslash}p{2.2cm}
  >{\centering\arraybackslash}p{3cm}
  >{\centering\arraybackslash}p{2.8cm}
  >{\centering\arraybackslash}p{3cm}
}
\toprule
\textbf{Reviewer ID} 
& \textbf{Documents} 
& \textbf{LLM vs Reviewer (Eligibility)} 
& \textbf{Reviewer Acceptance Rate} 
& \textbf{LLM vs Reviewer (Typology)} \\
\midrule
1           & 21  & 0.76 & 0.95 & 0.76 \\
\textbf{2}  & \textbf{216} & \textbf{0.85} & \textbf{0.96} & \textbf{0.68} \\
3           & 47  & 0.78 & 0.93 & 0.72 \\
4           & 23  & 0.82 & 0.95 & 0.73 \\
\textbf{5}  & \textbf{278} & \textbf{0.76} & \textbf{0.93} & \textbf{0.69} \\
6           & 17  & 0.58 & 0.64 & 0.58 \\
7           & 30  & 0.83 & 1.00 & 0.66 \\
\textbf{8}  & \textbf{132} & \textbf{0.71} & \textbf{0.97} & \textbf{0.75} \\
\bottomrule
\end{tabularx}
\caption{
Agreement between the LLM and human reviewers in filtering and classifying activities as either marketing or organizational. The \textit{Documents} column indicates the number of applications evaluated by each reviewer. \textit{Eligibility} measures agreement on whether activities qualify for funding. \textit{Acceptance Rate} reflects the proportion of initiatives accepted for funding. \textit{Typology} captures the agreement between the LLM and the reviewer when assigning the same category (marketing or organizational) to a given initiative.
}
\label{tab:reviewer_comparison_compete}
\end{table}

\begin{table}[h!]
\centering
\begin{tabularx}{0.7\textwidth}{
  >{\raggedright\arraybackslash}p{6cm}
  >{\centering\arraybackslash}X
  >{\centering\arraybackslash}X
}
\toprule
\textbf{Section} & \textbf{Cost (€)} & \textbf{Time (s)}\\
\midrule
Full summary             & 0.06 & 25 \\
Inovation activities     & 0.07 & 12 \\
Qualitative Analysis     & 0.03 & 13 \\
Inconsistency report     & 0.02 & 10 \\
Global score and rationale & 0.02 & 9 \\
\midrule
\textbf{Total}           & \textbf{0.20} & \textbf{69} \\
\bottomrule
\end{tabularx}
\caption{Cost and time analysis for \fambiental applications on average.}
\label{tab:time_cost_compete}
\end{table}

\newpage

\subsection{\compete Prompt Example}

\begin{table}[h!]
  \centering
  \begin{tabularx}{0.95\textwidth}{X}
    \toprule[1.1pt]

    \textbf{System Prompt:} A company is applying for funding with the goal of improving its operations. You are an evaluator of applications whose assessments will lead to either the approval or rejection of each funding proposal.\\ \\

    Write a project summary in European Portuguese (pt-PT), based only on the information provided about the company. Do not add external content. The summary should cover the following topics:\\ \\

    1. Business activity, recent developments, and relevant historical milestones;\\
    2. Main products and services and their relative weight in the business;\\
    3. Reference clients, sales structure (concentration/diversification);\\
    4. Export activity before the project;\\
    5. Other relevant attributes (Brands, Certifications, Awards/Insignia, Market positioning...);\\
    6. Main objectives underlying the funding proposal;\\
    7. Types of planned actions (an exhaustive list is not necessary);\\
    8. Export activity after the project (evolution of exports, market diversification, etc.)\\ \\

    The summary must always be structured in exactly five paragraphs, distributing the above topics logically among them.\\

\\

    \textbf{User Prompt:} This company is applying with an internationalization project. The relevant information provided by the company (which may contain minor errors), grouped by topic, is as follows:\\ \\

    General information about the beneficiary: \texttt{\{benef\_info\}}\\

    Brief summary of the proposed project: \texttt{\{cnd\_text\_info[6]\}}\\

    Company's areas of economic activity: \texttt{\{cnd\_text\_info[0]\}}\\

    Context, history, and evolution of the company: \texttt{\{cnd\_text\_info[1]\}}\\

    Currently relevant markets: \texttt{\{cnd\_text\_info[2]\}}\\

    New markets the company intends to reach: \texttt{\{cnd\_text\_info[3]\}}\\

    How the company intends to operate in these new markets: \texttt{\{cnd\_text\_info[4]\}}\\

    Extended summary of the project: \texttt{\{cnd\_text\_info[5]\}}\\

    Strategic diagnosis: \texttt{\{cnd\_text\_info[7]\}}\\

    Objectives associated with the investment: \texttt{\{cnd\_text\_info[8]\}}\\

    Technical description of the planned investments: \texttt{\{cnd\_text\_info[9]\}}\\

    Verified key market information: \texttt{\{merc\_text\}}\\

    \\
    \bottomrule[1.1pt]
  \end{tabularx}
  \caption{Example of the prompt used for generating a five paragraph application summary. The user prompt provides structured company and project data, using JSON query formatting. Some data, as the market information (stored in merc\_text), is pre-processed in order to reduce errors. A sixth paragraph, concerning financial values before and after the project is generated through logical inference based on the company’s financial sheets to avoid inaccuracies.}
\end{table}

\newpage
\subsection{\compete Reports Examples}

\subsubsection{Summary Report Example}
\begin{figure}[h!]
  \fbox{
  \includegraphics[width=\columnwidth, trim=50 175 30 50, clip]{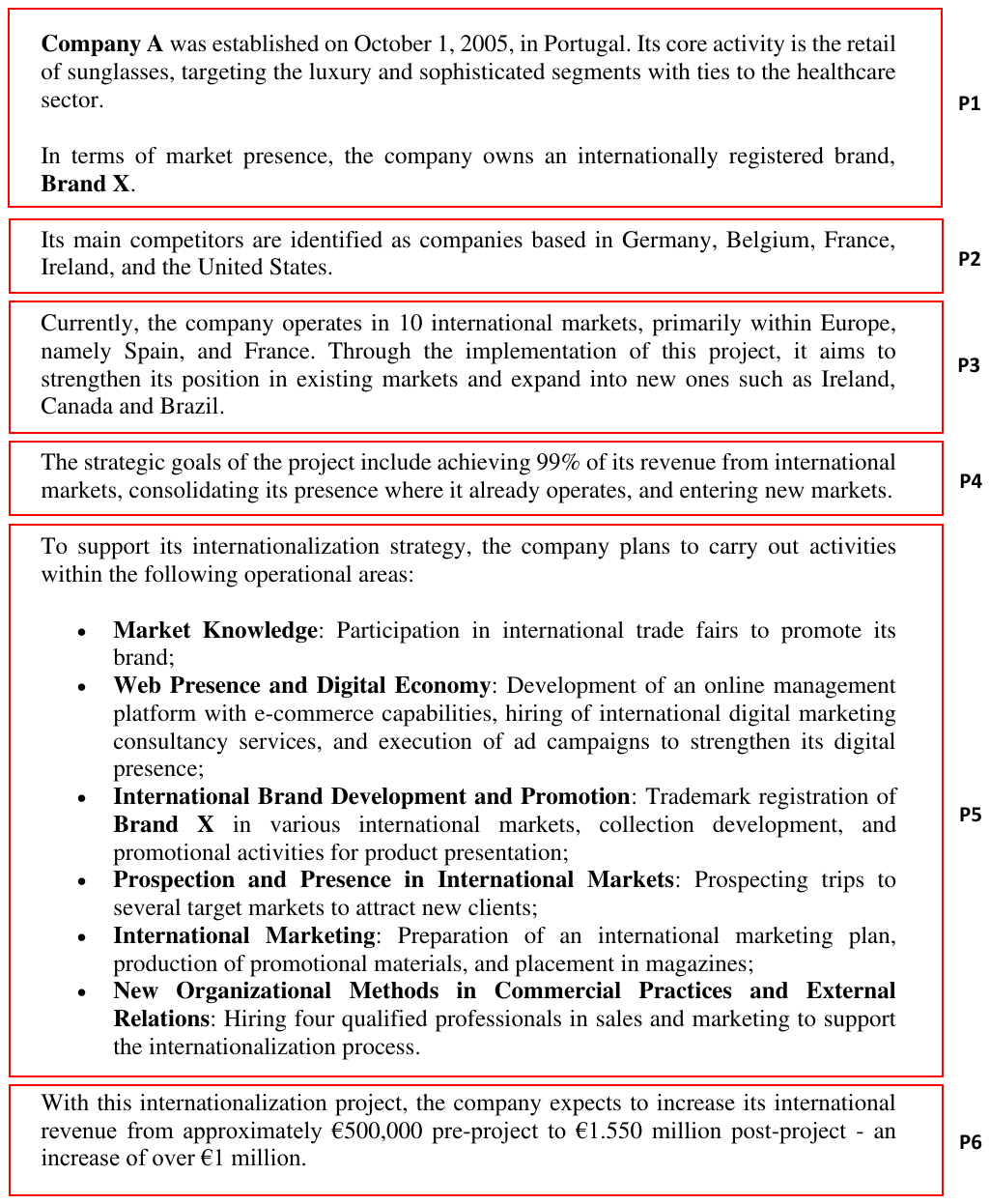}
  }
 \caption{\compete summary report example. The report is structured into six logical sections (referred to as paragraphs), each responsible for presenting a key component of the required information: P1 – Company Introduction, P2 – Market Activity, P3 – Competition, P4 – Strategic Objectives, P5 – Internationalization Activities, and P6 – Revenue before and after funding. Each paragraph is generated using structured JSON query formatting.}
  \label{fig:compete_report}
\end{figure}

\newpage 
\subsubsection{Qualitative Analysis Report Example}
\begin{figure}[h!]
  \fbox{
  \includegraphics[width=\columnwidth, trim=50 330 30 50, clip]{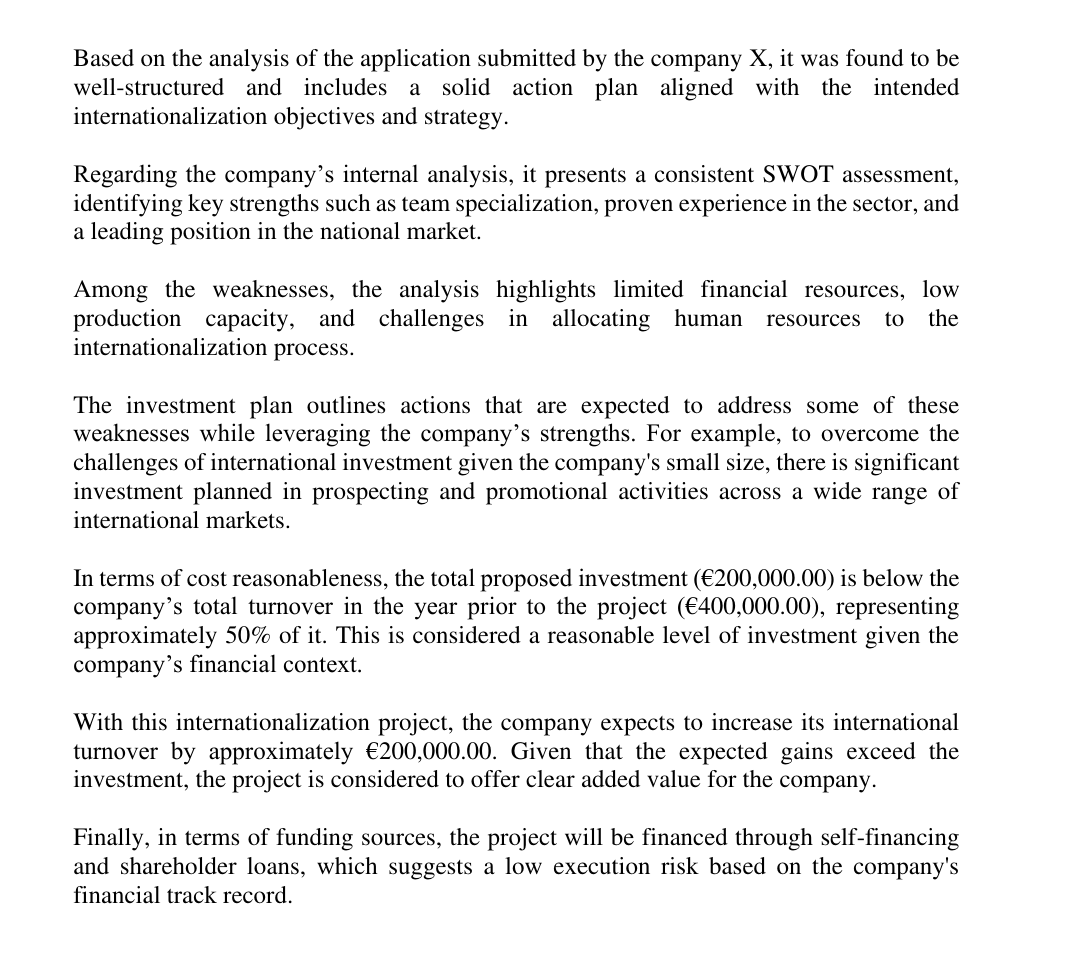}
  }
 \caption{\compete qualitative analysis report example. Each paragraph is generated using structured JSON query formatting.}
  \label{fig:compete_report}
\end{figure}

\newpage

\subsubsection{Innovation Activities Report Example}
\begin{figure}[h!]
  \fbox{
  \includegraphics[width=\columnwidth, trim=50 340 30 50, clip]{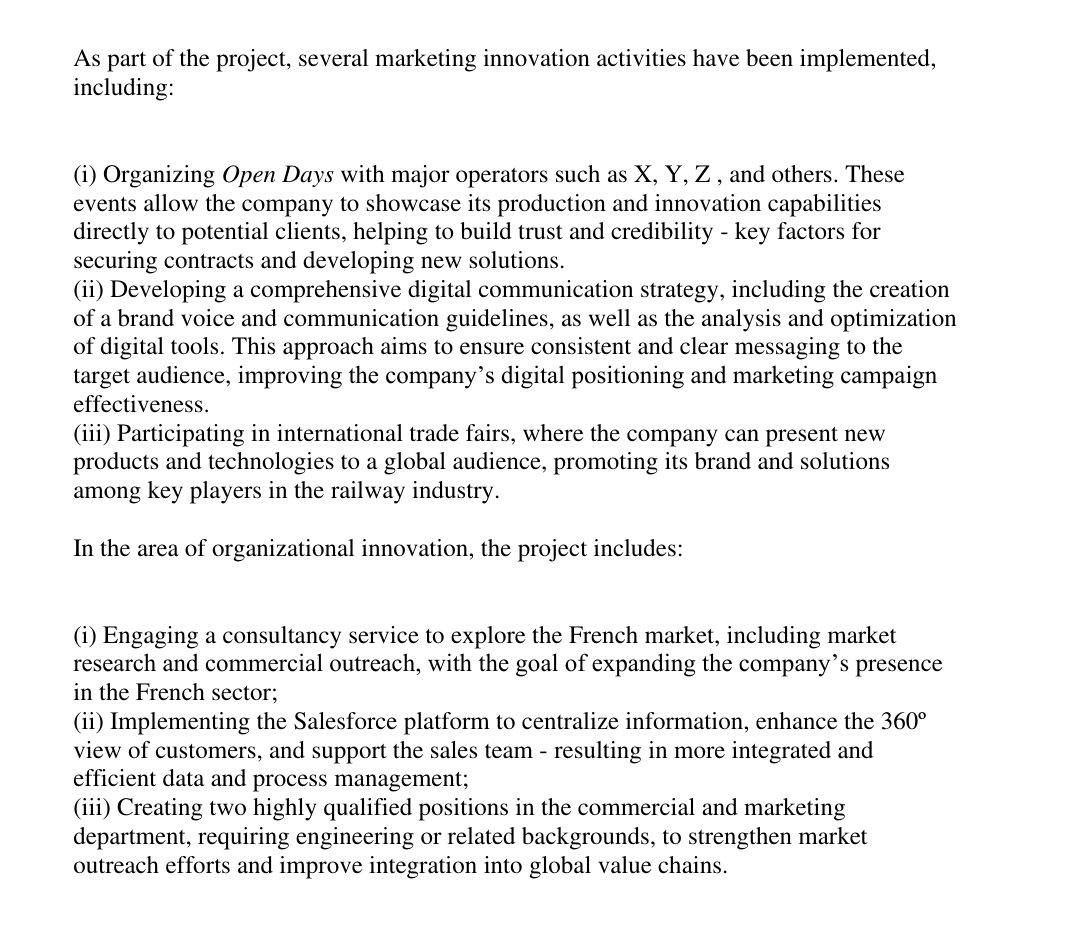}
  }
 \caption{\compete innovation activities report example.
The task involves selecting, from a list of proposed activities, those that are eligible, and classifying each as either organizational or marketing. Each paragraph is generated using structured JSON query formatting.}
  \label{fig:compete__inov_report}
\end{figure}